\documentclass[aps,pra,reprint,superscriptaddress]{revtex4-1}

\usepackage{physics}
\usepackage{graphicx}
\usepackage{amsmath,array,amssymb}
\usepackage{dcolumn}
\usepackage{bm}
\usepackage{hyperref}
\usepackage{color}

\graphicspath{{img/}}

\begin{document}
	
\title{Persistent atomic frequency comb based on Zeeman sub-levels of an erbium-doped crystal waveguide}

	\author{Mohsen Falamarzi Askarani}
	\affiliation{Institute for Quantum Science and Technology, and Department of Physics \& Astronomy, University of Calgary, 2500 University Drive NW, Calgary, Alberta, T2N 1N4, Canada}
	\affiliation{QuTech, Delft University of Technology, 2600 GA Delft, The Netherlands}
	
	\author{Thomas Lutz}
	\affiliation{Institute for Quantum Science and Technology, and Department of Physics \& Astronomy, University of Calgary, 2500 University Drive NW, Calgary, Alberta, T2N 1N4, Canada}
	\affiliation{ETH Zürich, Otto-Stern-Weg 1, 8093 Zürich, Switzerland}
	
	\author{Marcelli Grimau Puigibert}
		\affiliation{Institute for Quantum Science and Technology, and Department of Physics \& Astronomy, University of Calgary, 2500 University Drive NW, Calgary, Alberta, T2N 1N4, Canada}
	\affiliation{University of Basel, Klingelbergstrasse 82, CH-4056 Basel, Switzerland}
	
	\author{Neil Sinclair}
	\affiliation{Institute for Quantum Science and Technology, and Department of Physics \& Astronomy, University of Calgary, 2500 University Drive NW, Calgary, Alberta, T2N 1N4, Canada}
	\affiliation{Division of Physics, Mathematics and Astronomy and Alliance for Quantum Technologies (AQT), California Institute of Technology, 1200 East California Boulevard, Pasadena, California 91125, USA}
	
	\author{Daniel Oblak}
	\affiliation{Institute for Quantum Science and Technology, and Department of Physics \& Astronomy, University of Calgary, 2500 University Drive NW, Calgary, Alberta, T2N 1N4, Canada}

	\author{Wolfgang Tittel}
\affiliation{Institute for Quantum Science and Technology, and Department of Physics \& Astronomy, University of Calgary, 2500 University Drive NW, Calgary, Alberta, T2N 1N4, Canada}
\affiliation{QuTech, Delft University of Technology, 2600 GA Delft, The Netherlands}

	\date{\today}
	
\begin{abstract}
	Long-lived sub-levels of the electronic ground-state manifold of rare-earth ions in crystals can be used as atomic population reservoirs for photon echo-based quantum memories. We measure the dynamics of the Zeeman sub-levels of erbium ions that are doped into a lithium niobate waveguide, finding population lifetimes at cryogenic temperatures as long as seconds. Then, using these levels, we prepare and characterize  atomic frequency combs, which can serve as a memory for quantum light at 1532 nm wavelength. The results allow predicting a 0.1\%  memory efficiency, mainly limited by unwanted background absorption that we conjecture to be caused by the coupling between two-level systems (TLS) and erbium spins. Hence, while it should be possible to create an AFC-based quantum memory in Er$^{3+}$:Ti$^{3+}$:LiNbO$_3$, improved crystal growth together with optimized AFC preparation will be required to make it suitable for applications in quantum communication.
\end{abstract}
	
	
	\maketitle

\section{Introduction}
Cryogenically-cooled rare-earth-ion-doped (REI-doped) crystals have been extensively studied for their use in classical optical signal processing applications for several decades \cite{Tittel2010}. This is partially due to their convenient optical and spin-level structure, long population and coherence lifetimes, large inhomogeneous broadening, and their tunability with externally-applied fields \cite{liu2006sREIs-book}. More recently, this work has spawned applications in quantum signal processing, including photon echo-based quantum memories for light \cite{lvovsky2009optical,gundougan2015AFCspin-wave,NeilPLRmultiplexing,sabooni2013cavity,clausen2011quantumcrystal,saglamyurek2011broadband,AFC_spinwave}.

\begin{figure*}[t]
	\centering
	\includegraphics[width=2\columnwidth]{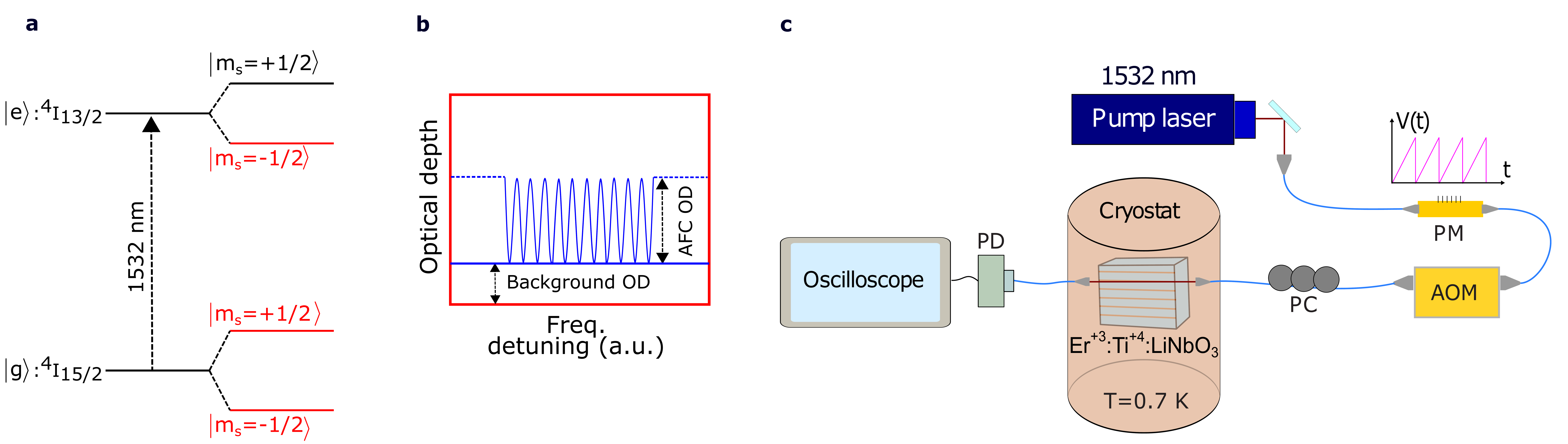}
\caption{\textbf{a.} Simplified energy level structure of the $^4$I$_{15/2}$ $\leftrightarrow$ $^4$I$_{13/2}$ transition of Er$^{3+}$. Exited and ground levels are indicated with $\ket{e}$ and $\ket{g}$, respectively, and electronic Zeeman sub-levels with $\ket{m_s=\pm 1/2}$. To create an atomic frequency comb (a periodic modulation of the frequency-dependent optical depth into equally-spaced narrow peaks), persistent spectral holes are created (burned) by pumping all undesired population into shelving levels, $ e.g. \ket{g, m_{s}=+1/2}$. \textbf{b.} An example of an AFC, showing the resulting spectral population grating. \textbf{c.} Experimental setup. Continuous-wave light at 1532 nm wavelength is directed through a phase-modulator (PM) and an acousto-optic modulator (AOM), which allow frequency and intensity modulation. After passing a polarization controller (PC), the light creates spectral holes and AFC structures in the erbium-doped lithium niobate waveguide, and furthermore allows probing previously created structures with the help of a photo-detector (PD) and an oscilloscope. The PM is driven by a serrodyne voltage V(t) modulation similar to that depicted in the plot.} 
	\label{fig:setup}
\end{figure*}
Several such protocols, including the widely-employed atomic frequency comb (AFC) protocol \cite{AFC}, require frequency-selective optical pumping (or persistent spectral hole burning) of REIs into a long-lived energy level, referred to as a shelving level (see Fig.\ref{fig:setup}a). Typically, this level is a spin (hyperfine) level within the electronic ground state manifold that features a much longer population lifetime than that of the optically-excited level. This allows waiting for excited atoms to decay at the end of the optical pumping sequence without loosing the spectral population grating in the ground state, which is key to quantum state storage with high fidelity and efficiency \cite{Spinmixing}.

Among the REIs, Er$^{3+}$ is the only one that features a transition from the ground level to an excited level at  telecommunication wavelength of around 1.5 $\mu$m. Since Er$^{3+}$ is a Kramer's ion, the ground level (more precisely, it's lowest-lying crystal field level) is split into two electronic Zeeman sub-levels under the application of a magnetic field. If a host crystal contains nuclear spins, these may couple to the Zeeman sub-levels of Er$^{3+}$, resulting in further splitting into superhyperfine levels \cite{liu2006sREIs-book}. Both Zeeman and superhyperfine levels are potentially useful as shelving levels for the AFC memory preparation, with the key requirement that they must feature a long population lifetime. But since the memory bandwidth assuming high-efficiency storage is limited by ground-state splitting, optical pumping into electronic Zeeman levels is preferred due to their larger splitting in a magnetic field.

Partially motivated by its remarkably long optical coherence lifetime of 4.4 ms \cite{4.4msT2YSO}, initial studies towards a telecommunication-wavelength quantum memory have focused on Er$^{3+}$:Y$_2$SiO$_{3}$ \cite{2010YSiOmemorynoise} and electronic Zeeman levels for spectral hole burning. However, their 130 ms population lifetime \cite{ZeemanYSO} (limited by  Er$^{3+}$ spin flip-flops \cite{bottger2006SD-Mechanisms}) in conjunction with 11 ms optical population lifetime have so far prevented reaching efficiencies in excess of 0.25 \%.
Later, AFC-based storage of entangled photons in an Er-doped SiO$_2$ fibre \cite{saglamyurek2015quantum} was achieved by exploiting spin disorder, which reduces spin flip-flops between Zeeman levels compared to those in crystals. On the other hand, the amorphous nature of SiO$_2$ also leads to small optical coherence lifetimes, thereby restricting storage times to less than 100 nanoseconds. Furthermore, the disorder-induced inhomogeneous broadening of the spin-transition has limited storage efficiencies to similar values as in Er$^{3+}$:Y$_2$SiO$_{3}$ \cite{hole-erbium,Coherence-fibre}. More recently, storage of heralded single photons using AFCs in an Er- and Ti-doped LiNbO$_3$ (Er$^{3+}$:Ti$^{3+}$:LiNbO$_3$) waveguide was achieved by taking advantage of population shelving in superhyperfine levels \cite{waveguide-storage}. However, as before, the efficiency did not exceed the percent level, this time due to remaining absorption in the AFC troughs caused by the complexity of the superhyperfine structure and excitation-induced spin relaxation \cite{waveguide-storage}. In addition, the superhyperfine splitting
limits the bandwidth for high-efficiency AFCs to around 100 MHz, even assuming magnetic fields of several Tesla. Finally, AFC-based storage of qubits encoded into attenuated laser pulses has been demonstrated using $^{167}$Er$^{3+}$:Y$_2$SiO$_{3}$. This work relied on spectral hole burning into nuclear spin levels of the $^{167}$Er isotope \cite{craiciu2019nanophotonic}, and the use of a nanocavity to reduce the lifetime of the excited level by means of the Purcell effect. However, the small storage bandwidth of 150 MHz, determined by the inhomogeneous linewidth of $^{167}$Er$^{3+}$:Y$_2$SiO$_{3}$, and the small efficiency of less than 1\% supports the general conclusion that the creation of a workable quantum memory for telecommunication-wavelength photons remains an open challenge. An interesting possibility is the use of  the same crystal in a magnetic fields of several Tesla \cite{ranvcic2018coherence}, but, so far, no storage experiment has been reported.

\begin{figure*}[t]
	\centering
	\includegraphics[width=\linewidth]{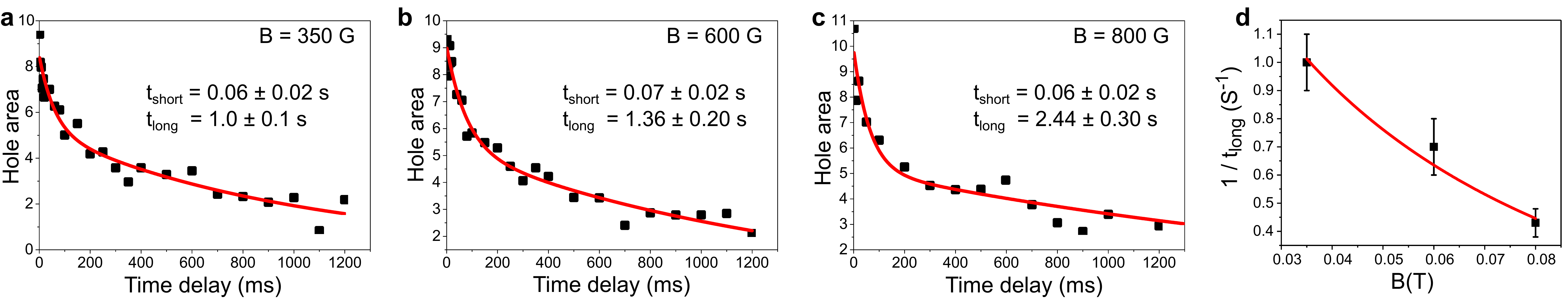}
	\caption{Time-resolved spectral hole decays at magnetic fields of \textbf{a.} 350 G, \textbf{b.} 600 G, and \textbf{c.} 800 G. \textbf{d.} Long-decay relaxation rate versus magnetic field.}
	\label{fig:lifetime}
\end{figure*}

Here we explore the generation of AFCs in Er$^{3+}$:Ti$^{3+}$:LiNbO$_3$ using Zeeman sub-levels for population shelving. First, we quantify the population lifetime of these levels using time-resolved spectral hole burning at a temperature of around 0.7~K and at magnetic fields of up to 1~kG. Next, we create AFCs that persist for up to a few seconds. However, the AFC structures are consistently marred by a significant absorption background, which restricts potential storage efficiencies.
Probing the origin of this background, we conjecture, after exclusion of other causes, that it is due to the coupling between laser-induced excitation of two-level systems (TLS) and erbium spins, leading to decay of the ground-state population grating. We conclude that improved crystal growth together with optimized AFC preparation will be required to create an efficient and high-bandwidth quantum memory.

\section{Experimental details}

Our experiments are performed using the $^4$I$_{15/2}$ $\leftrightarrow$ $^4$I$_{13/2}$ transition of Er$^{+3}$:Ti$^{4+}$:LiNbO$_{3}$, which is cooled to a temperature of around 0.7~K using an adiabatic demagnetization refrigerator. Details of the waveguide fabrication can be found in Ref. \cite{waveguide-storage}. The waveguide is exposed to magnetic fields up to 5 kG oriented parallel to the c-axis of the crystal. The field lifts the Kramer's degeneracy of the ground and excited electronic levels of erbium, giving rise to Zeeman levels (see Fig. \ref{fig:setup}a for a simplified energy level scheme of Er$^{3+}$). To interact with the erbium ions, we use light from a continuous-wave laser at around 1532 nm wavelength. As shown in Fig. \ref{fig:setup}c, it is frequency and intensity modulated, and then fibre butt-coupled into and out of the waveguide.

For time-resolved spectral hole burning measurements we first optically-excite erbium ions within a narrow spectral bandwidth, detuned by 250~MHz w.r.t the unmodulated laser light, using pulses of around 300 ms duration. The resulting decay redistributes the ions among the Zeeman sub-levels of the ground state, resulting in a spectral hole. 
After a varying time delay that exceeds the 2.1 ms population lifetime of the $^4$I$_{13/2}$ excited level, we measure the decay of the area of the spectral hole (which is proportional to the number of shelved atoms) by linearly varying the frequency of the laser light during 1 ms over a 100~MHz-wide frequency window surrounding the hole. Frequency sweeps are achieved by using a phase modulator and serrodyning, resulting in a modulation efficiency at 1~GHz detuning of approximately 50\%. 

For AFC generation, we frequency and intensity modulate the excitation light to burn up to 130 spectral pits over a bandwidth of up to 6.4 GHz during 300~ms. Each spectral pit is detuned by 50~MHz from its nearest neighbour to create a periodic modulation in the inhomogeneously broadened atomic absorption spectrum. This frequency spacing of the spectral pits corresponds to an AFC storage time of 20~ns \cite{AFC}. The absorption profile of the comb is read after a time delay of 30~ms by performing a frequency sweep over bandwidths of up to 6.4 GHz in 1 ms. All measurements are repeated 20 times, and hole and AFC absorption profiles are determined by averaging.

\section{Results and discussion}

\subsection{Population dynamics of ground state sub-levels} \label{section:spin inhom. extimation}

Time-resolved spectral hole burning is performed at fields of 350, 600, and 800 G. The time dependent decay of the hole area is plotted in Fig.~\ref{fig:lifetime}, indicating the occupation and the population lifetime of the ground-state Zeeman sub-levels along with those of any other shelving level \cite{hole-erbium}. We find that the best fit corresponds to a double exponential decay in which the shortest decay exhibits a $e^{-1}$ population lifetime of $t_\mathrm{short} \approx 0.06$~s that is independent of the magnetic field, whereas the $e^{-1}$ population lifetimes of the long decays are $t_\mathrm{long} = 1.00$, 1.36, and 2.44~s for magnetic fields of 350, 600, and 800~G, respectively. The relative weights of the exponentials do not change with field. Note that we are unable to burn the hole to transparency and we also do not resolve any side-holes or anti-holes, which arise from pumping of atomic population between different, well defined, atomic levels \cite{hastings2008spectral}.

The short, field independent decay is caused by population trapping in ground levels that couple only weakly to magnetic fields. Previous hole burning measurements \cite{waveguide-storage}, in conjunction with the temperatures and fields used here, suggest that these levels are likely superhyperfine levels arising from the coupling of Nb and Li spins of LiNbO$_3$. 

The approximately linear field dependence of the population lifetimes of the long decays ($t_{long}$) suggests that it is governed by spin flips \cite{hole-erbium}. An increasing magnetic field causes more ions to become spin polarized at low temperatures, which leads to a reduced spin flip-flop probability. 
Furthermore, the spin inhomogeneous broadening increases with field and reduces the flip-flop rate since it is less likely for two neighboring spins to be resonant \cite{hole-erbium}. 
We fit the relaxation rate of the long decays $1/t_{long}$ using a model that describes the temperature ($T$) and field ($B$) dependence of spin flip-flops and inhomogeneous broadening  \cite{hole-erbium}:
\begin{equation}
\frac{1}{t_{long}} = \frac{\alpha}{\Gamma_{s}+\gamma_{s} B} \,\, \mathrm{sech}^2 ( \frac{g\mu_{B}B}{2kT}).
\label{eq:spin flip-flops}
\end{equation}
The magnitude of the spin inhomogeneous broadening is described by a static term $\Gamma_{s}$ and a field-dependent term $\gamma_{s}B$, $g$ is the g-factor, $\mu_{B}$ the Bohr magneton, and $\alpha$ a scaling coefficient.

For a reliable fit of the limited experimental data shown in Fig. \ref{fig:lifetime}d, we assume $g=15.13$, which was inferred from measurements of an Er$^{3+}$:LiNbO$_3$ bulk crystal \cite{Charlesppt,thiel2010LNbulk}. Furthermore, fixing the scaling factor of $\alpha$ at 10$^9$~S$^{-2}$, as in \cite{hole-erbium}, we find static and field-dependent spin inhomogeneous broadenings of $\Gamma_{s}$=0.4$\pm$0.1~GHz and  $\gamma_{s}$=14.5$\pm$3.0~GHz/T, respectively. The relatively large values are not surprising given the significant spin inhomogeneous broadening of the nuclear Zeeman levels of Tm$^{3+}$ in a Ti$^{4+}$:LiNbO$_{3}$ waveguide \cite{sinclair2018}. These measurements imply significant magnetic disorder in Ti$^{4+}$:LiNbO$_{3}$, which is possibly due the inclusion of Ti$^{4+}$, near surface-related impurities, or the differences of congruent growth compared to other LiNbO$_{3}$ crystals \cite{sinclair2018}. 

\subsection{Creation of AFCs using Zeeman sub-levels}

\begin{figure}[t]
	\centering
	\includegraphics[width=0.7\columnwidth]{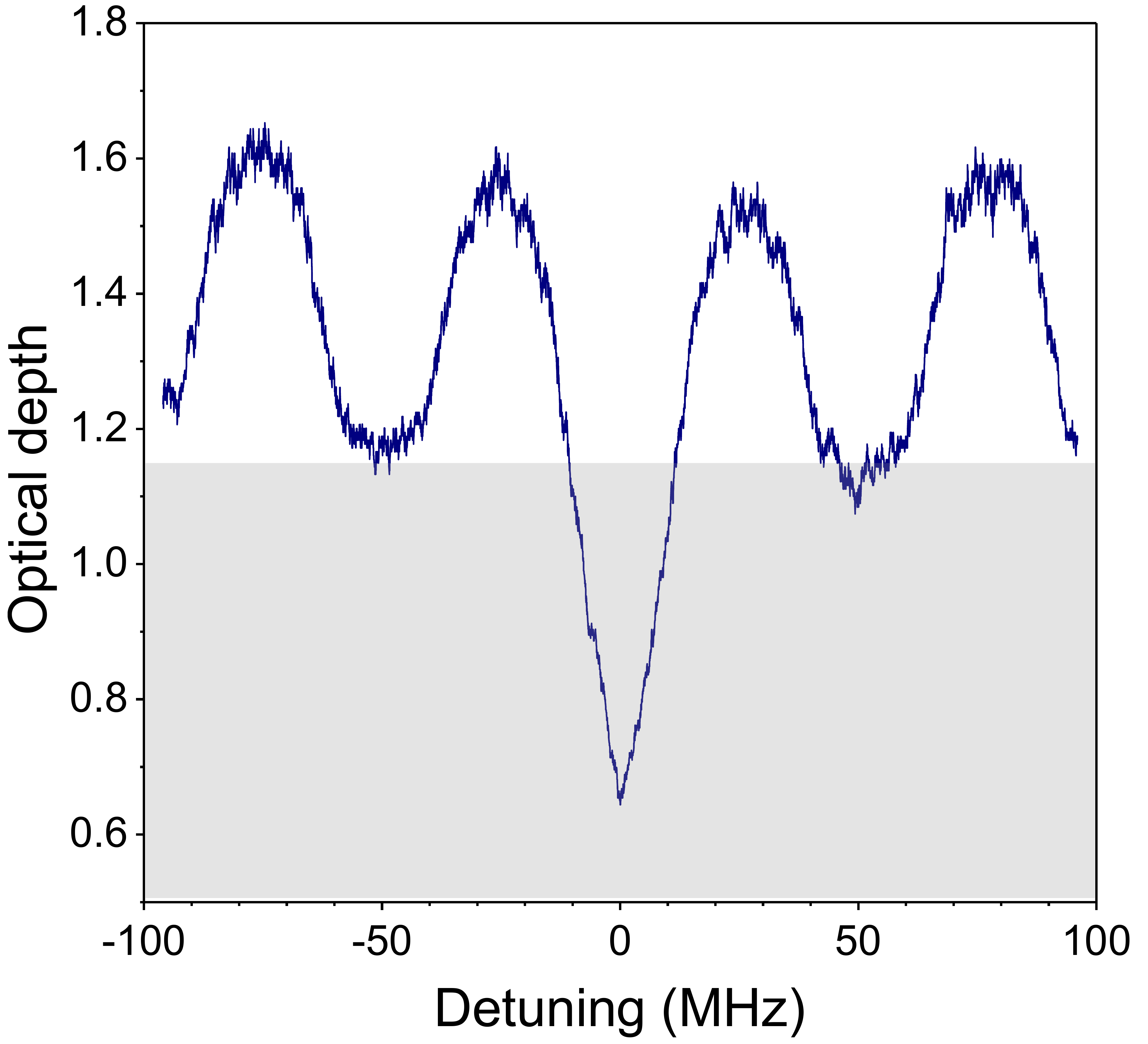}
	\caption{Absorption profile of 200 MHz-section of a 6.4 GHz-wide AFC at $\lambda$=1532.05~nm and B=3~kG. The grey shaded area indicates remaining background absorption.}
	\label{fig:Zeeman AFCs}
\end{figure}

\begin{figure}[t]
	\centering
	\includegraphics[width=0.8\columnwidth]{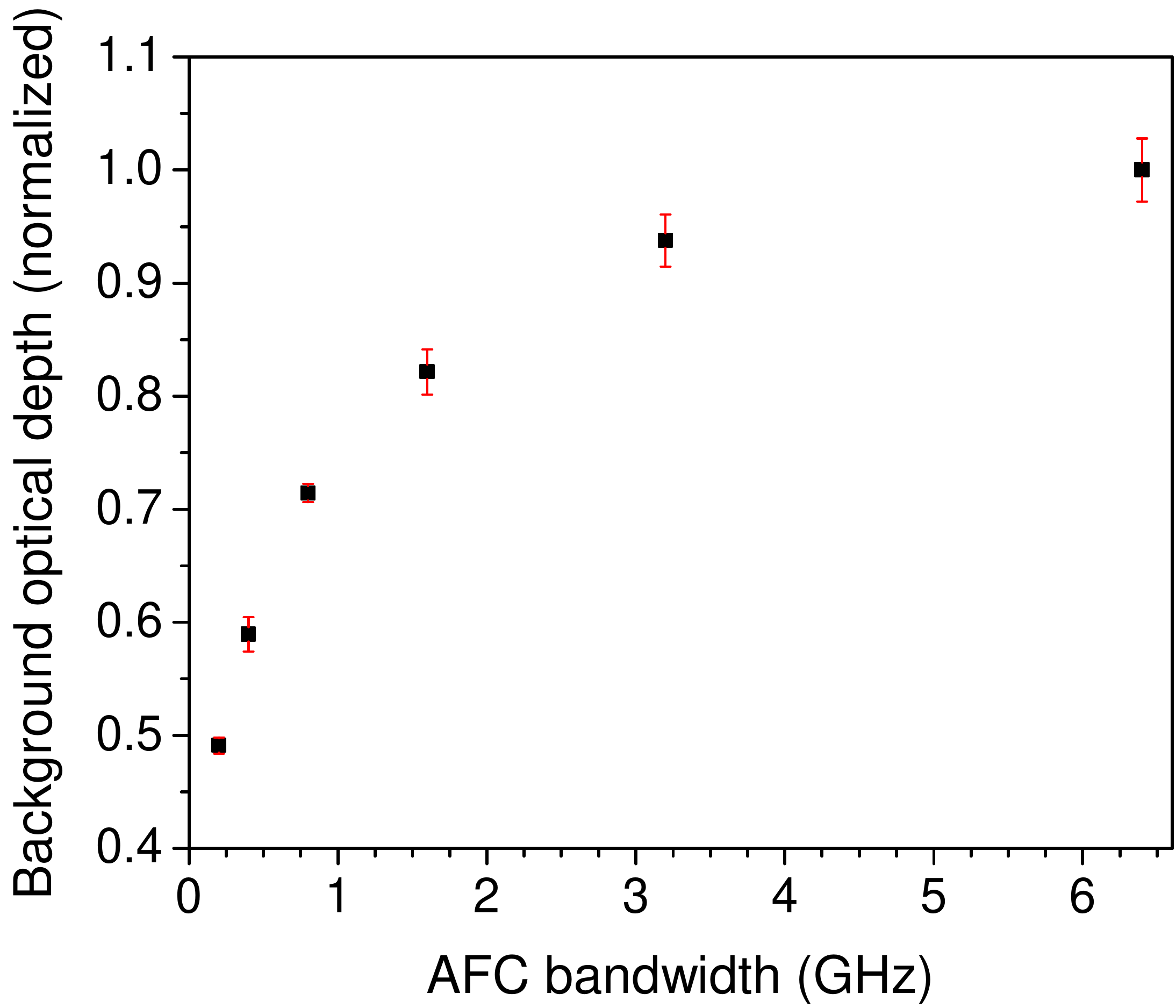}
	\caption{Average background absorption as a function of AFC bandwidth.}
	\label{fig:d0 vs AFC bandwidth (low OD)}
\end{figure}

In order to assess the possibility for broadband quantum memory using Zeeman sub-levels as population reservoir, we generate AFCs with bandwidths between 0.2 and 6.4~GHz symmetrically around zero detuning. All AFCs are centred at 1532.05 nm (corresponding to a spectral region with an optical depth of 2 for light propagating perpendicular to the crystal c-axis), are created under the application of a 3~kG magnetic field (oriented along the crystal c-axis), and feature peak spacings of 50~MHz. Furthermore, the overall duration of the optical pumping cycle is kept constant. Fig.~\ref{fig:Zeeman AFCs} shows a 200-MHz section of a 6.4-GHz wide AFC. Note the significant absorption background, which exponentially reduces the storage efficiency of our AFCs compared to the case of no background \cite{AFC}. The deep hole at zero detuning is due to optical pumping by unmodulated light leaking through the phase modulator.
As shown in Fig.~\ref{fig:d0 vs AFC bandwidth (low OD)}, we find that the background absorption, which is assessed at detunings between -100 and +100 MHz, increases when the bandwidth of the AFC increases. 

\subsection{Determining the origin of the background absorption}
\subsubsection{Anti-hole broadening}
One possible explanation for the bandwidth-dependent background absorption is that there is spectral overlap between broad anti-holes (caused by optical pumping) and the AFC. 
Provided the frequency difference between holes and anti-holes is larger than the anti-hole broadening, the spectral overlap---and hence the background absorption---increases as the AFC bandwidth approaches the hole-to-anti-hole splitting. This is the case in Er:LiNbO$_3$, for which the Zeeman level splitting at 3~kG  exceeds 50 GHz and the anti-hole broadening at the same field is expected to be only around 5~GHz.
However, the small AFC widths compared with the Zeeman splitting makes spectral overlap between AFCs and antiholes unlikely.

To fully rule out AFC background absorption due to inhomogeneous broadening of ground-state levels (whether arising from Zeeman splitting or not), we generate, at an optical depth of 0.8 (with light propagating parallel to the c-axis of the crystal), pairs of 200 MHz-bandwidth AFCs with varying detuning between their centre frequencies, up to 1.4~GHz. If the broadened anti-holes, created by the second AFC, indeed overlap with the the first AFC, we expect to see an increase in background optical depth of the first AFC. This back-filling method is similar to the one used for characterization of the inhomogeneous broadening of Zeeman sub-levels in an Er-doped fibre \cite{hole-erbium}. 
Table.~\ref{tab:anti hole effect} quantifies the background absorption of the first AFC at different detunings between the two AFCs. We find that measured values do not increase but rather scatter around a mean of around 0.66. This confirms that the anti-hole broadening does indeed not explain the observed increase of background shown in Fig.~\ref{fig:d0 vs AFC bandwidth (low OD)} -- at least not for AFC bandwidths up to around 1 GHz. Hence, another mechanism must be responsible, and we conjecture that it will also explain the background for large-bandwidths AFCs.

We note that the inhomogeneous broadening of the superhyperfine levels is too small at the applied magnetic fields to cause a constant absorption background \cite{waveguide-storage}, and their contribution to the observation in  Fig. \ref{fig:d0 vs AFC bandwidth (low OD)} can thus be ignored.

\begin{table}[htbp]
	\centering
	\caption{\bf Background absorption for 200 MHz-wide AFCs and varying frequency difference. The case of a single AFC (zero detuning) is included for reference.}
	\begin{tabular}{cc}
		\hline
		AFC & background\\ 
		detuning (GHz) & absorption \\
		\hline
		0 & 0.66 $\pm$ 0.01 \\
		0.6 & 0.60 $\pm$ 0.01 \\
		1.0 & 0.73 $\pm$ 0.03 \\
		1.4 & 0.66 $\pm$ 0.02\\
		\hline
	\end{tabular}
	\label{tab:anti hole effect}
\end{table}

\subsubsection{Instantaneous spectral diffusion (ISD)}
An alternative reason for the increase of background absorption with AFC bandwidth stems from the related increase of excited atoms during AFC preparation. 
(Even though the average laser power and pump-cycle-duration, i.e. the total energy of the pumping light, remain constant, the number of excited atoms grows due to the non-linear dependence of the absorption rate with the light power spectral density, which decreases with AFC bandwidth.)
This may result in two undesired processes: instantaneous spectral diffusion (ISD) \cite{thiel2014ISD} (discussed below) and erbium spin flips \cite{bottger2006SD-Mechanisms} (discussed in the following sections).  

Instantaneous spectral diffusion (ISD) can be introduced by optical pumping, during which Er$^{3+}$ ions are promoted to the excited state, leading to a uncontrollable shift of the transition energies of nearby Er$^{3+}$ ions compared to their unperturbed values. The consequence of addressing the latter ions during subsequent hole burning, is a smeared-out AFC with increased background. Indeed, as the cause for ISD---the presence of excited ions---will disappear at the end of the pumping sequence, transition frequencies will shift back to their unperturbed values, leading to a modification of the previously created absorption profile. ISD increases with the number of excited ions \cite{thiel2014ISD}, which is consistent with the observed increase of the background absorption with bandwidth. 

\begin{figure}[]
	\centering
	\includegraphics[width=0.9\columnwidth]{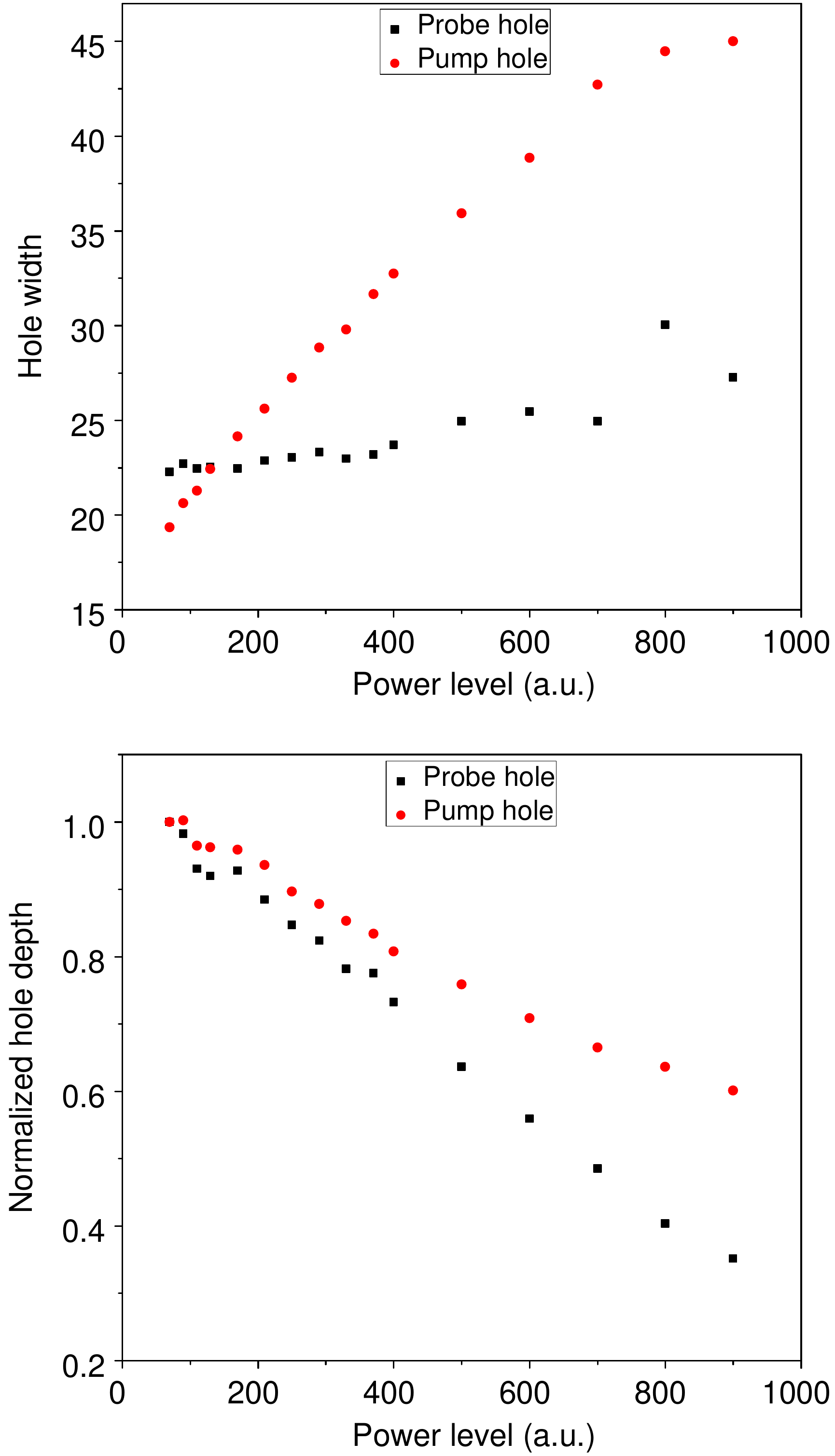}
	\caption{Laser excitation-induced change of a spectral hole. \textbf{a.} Widths and \textbf{b.} depths of the pump and probe holes (see text for details).}
	\label{fig:200-2hole}
\end{figure}

To characterise broadening and AFC background, which could potentially be due to ISD, we burn two 25 MHz-wide spectral holes with 200 MHz central frequency difference. We vary the laser excitation power used to create one of the holes, which we refer to as the pump hole, while constant power is used to burn the other, which we refer to as the probe hole. The width of the pump hole is expected to increase due to power broadening \cite{liu2006sREIs-book}, and ISD would manifest itself by simultaneous broadening and shallowing of the probe hole. We plot the widths and depths of the two holes as a function of optical excitation power used for the pump hole in Fig.~\ref{fig:200-2hole}a and b, and find that the width of the probe hole increases by around 5~MHz. Additionally, we observe decrease in the depths of the pump and probe holes with increasing pump power (Fig.~\ref{fig:200-2hole}b). Note that due to the optical pumping the reduction of the pump hole depth with increasing pump power is less than that of the probe hole, which decreases by factor of three. The shallowing of the probe and pump holes is consistent with the increase in the background absorption of the AFCs with their bandwidths as shown in Fig. \ref{fig:d0 vs AFC bandwidth (low OD)}.
However, the level of probe hole broadening of only 5~MHz neither explains the large shallowing of the probe hole nor the increased background optical depth observed in AFCs with 50 MHz tooth spacing (although it will impact the quality of AFCs with small peak spacing, i.e. AFCs that allow for longer storage times).

We also estimate the effect of ISD from measurements of a 0.1\%Er$^{3+}$:LiNbO$_3$ bulk crystal \cite{Charlesppt}. Assuming a maximum laser excitation power of $\sim$0.5~mW, 0.2\% Er$^{3+}$ concentration, and an ISD coefficient of $\sim$2$\times$10$^{-13}$ Hz $\cdot$ cm$^{3}$/excited ion, we predict the maximum spectral broadening to be on the order of kHz. This value is three orders of magnitude smaller than the observed increase of the width of the probe hole in Fig. \ref{fig:200-2hole}a. Hence, our experimental results and our estimate suggest that ISD is neither the cause for the observed spectral broadening of the probe hole, nor for the remaining background absorption of the probe hole and the AFCs.
In the next sections we elaborate on two other power-dependent mechanisms that may cause these observations.

\subsubsection{Spin flip-flops}\label{section:flip-flops}

Another process that could be affected by the increase of Er$^{3+}$ excitation when creating AFCs of larger bandwidths is the number of spin flip-flops. In this case, spins in one ground-state sub-level (Zeeman shelving level) that are within the AFC bandwidth resonantly exchange state with other spins that are in the other ground-state sub-level and outside of the AFC bandwidth: one spin will be flipped up, the other flopped down. This leads to redistribution of the population within the AFC, and hence to background. However, as the spin flip-flop rate decreases with the sixth power of the distance between ER-ions, we estimate its contribution to be small. In fact, if we extrapolate results from \cite{petersen2017Rateflip-flop} to an approximate Er-doping concentration of 3.6$\times$10$^{19}$~cm$^{-3}$ (or 0.2\%), we find a flip-flop rate of a few Hz -- by far to little to explain the AFC background. \\

\subsubsection{TLS-driven spin-flips}\label{section:TLS-flips}

Another power-dependent mechanism that may be responsible for the increase of the width of the probe hole and the AFC background is light-induced two-level system (TLS) excitation, which, in turn, may drive spin flips. In this case, the laser field first excites TLSs in LiNbO$_3$ \cite{TLShole-filling}. The TLSs may subsequently decay into lower-energy states and emit phonons, which interact with the electronic spins of Er$^{3+}$, leading to spectral diffusion that can explain the observations in Fig. \ref{fig:d0 vs AFC bandwidth (low OD)}. We note that a similar effect was observed in Tm$^{3+}$-doped Ti$^{3+}$:LiNbO$_3$ \cite{sinclair2018}.

In support of this hypothesis, we note that congruent LiNbO$_3$ is known to contain various impurities and imperfections, and previous measurements of REI-doped LiNbO$_3$ suggest significant lattice disorder. Such disorder could be the origin of the above-described TLSs \cite{sinclair2018,Charles2012Tm:LN,thiel2010LNbulk,2004reviewLNcrsytals}. We also point out that laser-induced excitation and decay of TLS impurities determines photo-refraction in LiNbO$_3$. It is well-known that this effect is enhanced in waveguides compared to bulk crystals due to high confinement of the laser field and, consequently, larger light intensities. This is consistent with the fact that the emergence of an absorption background in AFCs due to spectral hole filling is not observed in bulk LiNbO$_3$. 

\subsection{Discussion and conclusion}

We experimentally study ground-state Zeeman sub-levels of Er$^{3+}$ ions doped into a Ti$^{3+}$:LiNbO$_3$ crystalline waveguide as shelving levels for AFC-type quantum memory for light. Despite promising lifetimes in excess of a second, several changes are required to increase the storage efficiency for telecommunication-wavelength photons beyond the current state-of-the-art of around 1\%. Most importantly, the memory efficiency, which is limited by the available optical depth (at most 2 in our waveguide) and the absorption background need to be improved. The former can be addressed using an impedance-matched cavity \cite{unitefficiency2}, and the absorption background may be reduced through optimization of the optical pumping procedure. This includes the use of optimized optical excitation power, laser scan rate, and magnetic field strength and direction, which may result in longer-lived Zeeman levels and a favourable branching ratio into the shelving level. But ultimately, it seems necessary to improve the LiNbO$_3$ crystal itself, e.g. by varying growth conditions, to reduce the number of two-level systems. 

\section{Acknowledgments}
The authors thank Wolfgang Sohler, Mathew George, and Raimund Ricken for providing the waveguide, and Jacob H. Davidson for help with aligning the waveguide and Gustavo Amaral and Erhan Saglamyurek for useful discussions. The authors also acknowledge support from Alberta Innovates (AI), the Alberta Major Innovation Fund, the Natural Sciences and Engineering and Research Council of Canada (NSERC), and the Dutch Organisation for Scientific Research (NWO). N.S. acknowledges funding from the AQT's Intelligent Quantum Networks and Technologies (INQNET) research program, and W.T. support as a Senior Fellow of the Canadian Institute for Advanced Research (CIFAR).


\begin{thebibliography}{10}
	\newcommand{\enquote}[1]{``#1''}
	
	
	\bibitem{Tittel2010}
	W.~Tittel, M.~Afzelius, T.~Chanelière, R.~Cone, S.~Kr\"oll, S.~Moiseev, M.~Sellars, \emph{Laser and Photonics Review}, \textbf{4} , 244-267 (2010).
	
	\bibitem{liu2006sREIs-book}
	G.~Liu and B.~Jacquier, \emph{Spectroscopic properties of rare earths in
		optical materials}, vol.~83 (Springer Science \& Business Media, 2006).
	
	\bibitem{lvovsky2009optical}
	A.~I. Lvovsky, B.~C. Sanders, and W.~Tittel, \enquote{Optical quantum memory,}
	Nature photonics \textbf{3}, 706--714 (2009).
	
	\bibitem{gundougan2015AFCspin-wave}
	M.~G{\"u}ndo{\u{g}}an, P.~M. Ledingham, K.~Kutluer, M.~Mazzera, and
	H.~de~Riedmatten, \enquote{Solid state spin-wave quantum memory for time-bin
		qubits,} Physical review letters \textbf{114}, 230501 (2015).
	
	\bibitem{NeilPLRmultiplexing}
	N.~Sinclair, E.~Saglamyurek, H.~Mallahzadeh, J.~A. Slater, M.~George,
	R.~Ricken, M.~P. Hedges, D.~Oblak, C.~Simon, W.~Sohler, and W.~Tittel,
	\enquote{Spectral multiplexing for scalable quantum photonics using an atomic
		frequency comb quantum memory and feed-forward control,} Phys. Rev. Lett.
	\textbf{113}, 053603 (2014).
	
	\bibitem{sabooni2013cavity}
	M.~Sabooni, S.~T. Kometa, A.~Thuresson, S.~Kr{\"o}ll, and L.~Rippe,
	\enquote{Cavity-enhanced storage--preparing for high-efficiency quantum
		memories,} New Journal of Physics \textbf{15}, 035025 (2013).
	
	\bibitem{clausen2011quantumcrystal}
	C.~Clausen, I.~Usmani, F.~Bussieres, N.~Sangouard, M.~Afzelius,
	H.~de~Riedmatten, and N.~Gisin, \enquote{Quantum storage of photonic
		entanglement in a crystal,} Nature \textbf{469}, 508--511 (2011).
	
	\bibitem{saglamyurek2011broadband}
	E.~Saglamyurek, N.~Sinclair, J.~Jin, J.~A. Slater, D.~Oblak, F.~Bussieres,
	M.~George, R.~Ricken, W.~Sohler, and W.~Tittel, \enquote{Broadband waveguide
		quantum memory for entangled photons,} Nature \textbf{469}, 512--515 (2011).
	
	\bibitem{AFC_spinwave}
	M.~Afzelius, I.~Usmani, A.~Amari, B.~Lauritzen, A.~Walther, C.~Simon,
	N.~Sangouard, J.~Min{\'a}{\v{r}}, H.~de~Riedmatten, N.~Gisin, and S.~Kr\"oll,
	\enquote{Demonstration of atomic frequency comb memory for light with
		spin-wave storage,} Phys. Rev. Lett. \textbf{104}, 040503 (2010).
	
	\bibitem{AFC}
	M.~Afzelius, C.~Simon, H.~de~Riedmatten, and N.~Gisin, \enquote{Multimode
		quantum memory based on atomic frequency combs,} Phys. Rev. A \textbf{79},
	052329 (2009).
	
	
	\bibitem{Spinmixing}
	B.~Lauritzen, S.~Hastings-Simon, H.~De Riedmatten, M.~Afzelius, and N.~Gisin, \enquote{State preparation by optical pumping in erbium-doped solids using stimulated emission and spin mixin,} Phys. Rev. A \textbf{78},
	043402 (2008). 
	
	
	\bibitem{4.4msT2YSO}
	T.~B\"ottger, C.~W. Thiel, R.~L. Cone, and Y.~Sun, \enquote{Effects of magnetic
		field orientation on optical decoherence in
		{E}r$^{3+}$:{Y}$_{2}${S}i{O}$_{5}$,} Phys. Rev. B \textbf{79}, 115104 (2009). 
	
	\bibitem{2010YSiOmemorynoise}
	B.~Lauritzen, J.~Min{\'a}{\v{r}}, H.~de~Riedmatten, M.~Afzelius, N.~Sangouard,
	C.~Simon, and N.~Gisin, \enquote{Telecommunication-wavelength solid-state
		memory at the single photon level,} Phys. Rev. Lett. \textbf{104}, 080502
	(2010).
	
	\bibitem{ZeemanYSO}
	S.~Hastings-Simon, B.~Lauritzen, M.~U. Staudt, J.~L.~M. van Mechelen, C.~Simon,
	H.~De~Riedmatten, M.~Afzelius, and N.~Gisin, \enquote{Zeeman-level lifetimes
		in {E}r$^{3+}$:{Y}$_{2}${S}i{O}$_{5}$,} Physical Review B \textbf{78}, 085410
	(2008).
	
	\bibitem{bottger2006SD-Mechanisms}
	T.~B{\"o}ttger, C.~Thiel, Y.~Sun, and R.~Cone, \enquote{Optical decoherence and
		spectral diffusion at 1.5 $\mu$ m in {E}r$^{+3}$:{Y}$_{2}${S}i{O}$_{5}$
		versus magnetic field, temperature, and {E}r$^{+3}$ concentration,} Physical
	Review B \textbf{73}, 075101 (2006).
	
	\bibitem{saglamyurek2015quantum}
	E.~Saglamyurek, J.~Jin, V.~B. Verma, M.~D. Shaw, F.~Marsili, S.~W. Nam,
	D.~Oblak, and W.~Tittel, \enquote{Quantum storage of entangled
		telecom-wavelength photons in an erbium-doped optical fibre,} Nature
	Photonics \textbf{9}, 83--87 (2015).
	
	\bibitem{hole-erbium}
	E.~Saglamyurek, T.~Lutz, L.~Veissier, M.~P. Hedges, C.~W. Thiel, R.~L. Cone,
	and W.~Tittel, \enquote{Efficient and long-lived zeeman-sublevel atomic
		population storage in an erbium-doped glass fibre,} Phys. Rev. B \textbf{92},
	241111 (2015).
	
	\bibitem{Coherence-fibre}
	L.~Veissier, M.~Falamarzi, T.~Lutz, E.~Saglamyurek, C.~W. Thiel, R.~L. Cone,
	and W.~Tittel, \enquote{Optical decoherence and spectral diffusion in an
		erbium-doped silica glass fibre featuring long-lived spin sublevels,} Phys. Rev. B \textbf{94}, 195138 (2016).
	
	\bibitem{waveguide-storage}
	M.~F. Askarani, M.~Pugibert, T.~Lutz, V.~B. Verma, M.~D. Shaw, S.~W. Nam,
	N.~Sinclair, D.~Oblak, and \emph{et~al.}, \enquote{Storage and re-emission of
		heralded telecommunication-wavelength single photons using a crystal
		waveguide,} \textbf{11}, 054056 Phys. Rev. Applied (2019).
	
	\bibitem{craiciu2019nanophotonic}
	I.~Craiciu, M.~Lei, J.~Rochman, J.~M. Kindem, J.~G. Bartholomew, E.~Miyazono,
	T.~Zhong, N.~Sinclair, and A.~Faraon, \enquote{Nanophotonic quantum storage
		at telecommunications wavelength,} arXiv preprint arXiv:1904.08052  (2019).
	
	\bibitem{ranvcic2018coherence}
	M.~Ran{\v{c}}i{\'c}, M. P.~Hedges, R. L.~Ahlefeldt, and M.~J. Sellars, \enquote{Coherence time of over a second in a telecom-compatible quantum memory storage material,} Nature Physics \textbf{14}, 50 (2018).
	
	\bibitem{hastings2008spectral}
	S.~Hastings-Simon, M.~Afzelius,  J.~Min{\'a}{\v{r}}, M. U.~Staudt, B.~Lauritzen, H.~ de Riedmatten, N.~Gisin, A.~Amari, A.~Walther, S.~Kr{\"o}ll, and \emph{et~al.}
	\enquote{Spectral hole-burning spectroscopy in {Nd}$^{3+}$:{YVO}$_{4}$,} Phys. Rev. B \textbf{77}, 125111 (2008).
	
	\bibitem{Charlesppt}
	C.~W. Thiel, R.~M. Macfarlane, R.~L. Cone, Y.~Sun, T.~Bottger, K.~D. Merkel,
	and B.~W. R, \enquote{Spectroscopy and dynamics of
		{E}r$^{+3}$:{L}i{N}b{O}$_{3}$ at 1.5 microns for quantum information and
		signal processing applications,}  (10th Intl Meeting on Hole Burning, Single
	Molecule, and Related Spectroscopies: Science and Applications - HBSM 2009).
	
	\bibitem{thiel2010LNbulk}
	C.~Thiel, R.~Macfarlane, T.~B{\"o}ttger, Y.~Sun, R.~Cone, and W.~Babbitt,
	\enquote{Optical decoherence and persistent spectral hole burning in
		{E}r$^{3+}$: {L}i{N}b{O}$_{3}$,} Journal of Luminescence \textbf{130},
	1603--1609 (2010).
	
	\bibitem{sinclair2018}
	N.~Sinclair, C.~W. Thiel, D.~Oblak, E.~Saglamyurek, R.~L. Cone, and W.~Tittel,
	\enquote{Properties of a {T}m-doped {L}i{N}b{O}$_{3}$ waveguide at low
		temperatures,} In preparation  (2018).
	
	\bibitem{thiel2014ISD}
	C.~W. Thiel, R.~M. Macfarlane, Y.~Sun, T.~B{\"o}ttger, N.~Sinclair, W.~Tittel,
	and R.~L. Cone, \enquote{Measuring and analyzing excitation-induced
		decoherence in rare-earth-doped optical materials,} Laser Physics
	\textbf{24}, 106002 (2014).
	
	\bibitem{petersen2017Rateflip-flop}
	E.~S. Petersen, A.~M. Tyryshkin, K.~M. Itoh, M.~L. Thewalt, and S.~A. Lyon,
	\enquote{Measuring electron spin flip-flops through nuclear spin echo
		decays,} arXiv preprint arXiv:1709.02881  (2017).
	
	\bibitem{TLShole-filling}
	R.~Jankowiak and G.~Small, \enquote{Hole-burning spectroscopy and relaxation
		dynamics of amorphous solids at low temperatures,} Science \textbf{237},
	618--625 (1987).
	
	\bibitem{Charles2012Tm:LN}
	Y.~Sun, C.~Thiel, and R.~Cone, \enquote{Optical decoherence and energy level
		structure of 0.1\% {T}m$^{+3}$: {L}i{N}b{O}$_{3}$,} Physical Review B
	\textbf{85}, 165106 (2012).
	
	
	\bibitem{2004reviewLNcrsytals}
	L.~Arizmendi, \enquote{Photonic applications of lithium niobate crystals,}
	physica status solidi (a) \textbf{201}, 253--283 (2004).
	
	\bibitem{unitefficiency2}
	M.~Afzelius and C.~Simon,
	\enquote{Impedance-matched cavity quantum memory,} Phys. Rev. A \textbf{82},
	022310 (2010).
	
\end{thebibliography}
\end{document}